# String theory, Einstein, and the identity of physics: Theory assessment in absence of the empirical

Jeroen van Dongen

*Institute for Theoretical Physics*
*Vossius Center for History of the Humanities and Sciences*
*University of Amsterdam, Amsterdam, The Netherlands*

*Abstract*

String theorists are certain that they are practicing physicists. Yet, some of their recent critics deny this. This paper argues that this conflict is really about who holds authority in making rational judgment in theoretical physics. At bottom, the conflict centers on the question: *who is a proper physicist?* To illustrate and understand the differing opinions about proper practice and identity, we discuss different appreciations of epistemic virtues and explanation among string theorists and their critics, and how these have been sourced in accounts of Einstein's biography. Just as Einstein is claimed by both sides, historiography offers examples of both successful and unsuccessful non-empirical science. History of science also teaches that times of conflict are often times of innovation, in which novel scholarly identities may come into being. At the same time, since the contributions of Thomas Kuhn historians have developed a critical attitude towards formal attempts and methodological recipes for epistemic demarcation and justification of scientific practice. These are now, however, being considered in the debate on non-empirical physics.

**Introduction**

Theoretical high energy physics is in crisis. Many physicists may wish to deny this, but it is richly illustrated by the heated exchanges, charged manifestos and exclamations of despair in highly visible publications. For example, three prominent cosmologists, Anna Ijjas, Paul Steinhardt and Abraham Loeb, argued in the February 2017 issue of *Scientific American* that the long favoured model for the early universe, inflationary cosmology, has no data to support it and has gone through so many patch-ups that it is now beyond testability. In fact, the authors contended, those still favouring it are squarely, by explicit choice or implicit commitment, marching towards a fundamentally "non-empirical science" (Ijjas et al., 2017, p. 39).

The charge has met with a strong rebuke: Alan Guth, one of inflation's inventors, together with thirty-two equally prominent other signatories (including five Nobel laureates) expressed "categorical disagreement" with the *Scientific American* article (Guth *et al.* 2017). Emphasizing that they represent the consensus point of view, they are "bewildered" by the criticism: inflationary cosmology is not about any one particular model, but a program, a "class of models based on similar principles" that has actually achieved "impressive empirical success": some of inflation's key predictions, such as particular features of the cosmic microwave background radiation, are argued to have been convincingly confirmed. Thus,

Guth *et al*. reject the claim that the inflation scenario is untestable and in any case insist that "the testability of a theory in no way requires that all its predictions be independent of the choice of parameters." Even if it is "standard practice in empirical science to modify a theory as new data come to light", as they argue was done in the construction of the Standard Model of particle physics, Guth and his co-authors dispute that inflation has been patched: there has been "no need to go beyond the class of standard inflationary models." Speaking as conscientious empiricist cosmologists, they insist that "empirical science is alive and well!"

Such heated exchanges have not been limited to cosmology. *The New York Times* reported in June 2017 that researchers at CERN are increasingly nervous about the lack of data on the existence of supersymmetry (Overbye 2017), a key element of string theory, which has not yet shown itself in any of CERN's experiments. Many related searches for dark matter particles—expected to be necessary to plug gravity shortages in our cosmologies—have also come up empty handed (Bertone and Tait 2018). "Many of my colleagues are desperate", German theoretical physicist Hermann Nicolai was quoted in the *New York Times* article. The flexibility of theory is considered to be particularly problematic: the parameters of supersymmetry, like the models of dark matter, have been adjusted a number of times so that there always seems to be another version that might yet show itself in yet another attempt at observation—so far, however, it has not. Particle physicist Sabine Hossenfelder has been most explicit in expressing despair, for instance in the pages of *Nature Physics*: according to her, fundamental physics is facing a dearth of data and a bewildering plethora of theories, multiplying primarily by what attracts citations and funding. "The current practices in theory development signal a failure of the scientific method", Hossenfelder believes. She contends that, while astrophysicists propose more and more candidates for dark matter or different models for inflation, "in the absence of good quality measures [...] there is no evidence that a theory's fruitfulness correlates with its correctness." While fashion dictates, objectivity is impaired: "social and cognitive biases" have produced "a serious systemic failure."[1]

The carefully muted excitement over the most recent preliminary announcements coming from CERN and Fermilab of possible physics beyond the Standard Model illustrates both the current apprehension as much as cheer for a promise of a resolution of the dearth of data. While results are still inconclusive, there are hopes of eventually confirming novel leptoquarks or an anomalous magnetic moment for the muon. Yet, the interpretation of Fermilab's data for the magnetic moment are still debated, as much as they are only preliminary, while CERN is working to improve its error bars in its meson decay measurements.[2] One way or another, the announcement of the results, as experimentalist Tomasso Dorigo put it, are "good for particle physics [...] because particle physics has been dead for a while." Hossenfelder, in her somewhat characteristic acerbic fashion, "wouldn't bet" on any new physics materializing.[3] Yet, more to the point: it is hardly to be expected that these results, if in the end confirmed, will reach directly into the depths of quantum gravity, where string theorists traditionally do their work.

---

[1] Hossenfelder (2017), p. 316; see also Hossenfelder (2018); Butterfield (2019).
[2] For more on Fermilab's results, see e.g. Overbye (2021); Abi et al. (2021); for CERN, see Sample (2021); De Vries (2021); and the LHC seminars of 23 March 2021 at https://indico.cern.ch/event/976688/.
[3] Dorigo as in Garisto (2021); Hossenfelder is quoted in Overbye (2021).

Indeed, the crisis in fundamental physics has manifested itself most markedly in discussions on string theory. The subject, however, has remained equally influential as it is ambitious. Hossenfelder accuses string theory of a particular disregard for "facts", which she believes mirrors the familiar disregard in recent political discourse.[4] Yet high-energy theorists Herman Verlinde and Nima Arkhani-Hamed, both in Princeton, paint a strikingly different picture: while they equally lament post-factual politics, they put forward their own sub-discipline as the bulwark against this alarming trend: fundamental physics and string theory are all about, in their words, finding the "Truth with capital T." The pursuit of these subjects would discipline their students into "tolerance, stamina, reflection and the ability to question your assumptions."[5] Still, string theory's critics point especially to its cosmologies as stark examples of fact-free physics.

Fundamental high energy physics does not have a tradition of public conflict. Midway through the twentieth century it was the first subject to see large-scale Big Science collaboration teams. These, even if fiercely competing with one another, internally upheld a cultural norm that valued conformity and team spirit, and suppressed public dissent.[6] Its theoretical sub-discipline, too, was once an outwardly uniform field in which philosophical debate was either absent or quickly stifled.[7] Yet, the above suggests that this sub-discipline is now becoming increasingly frayed: the disputes on what counts as solid method suggest a *breakdown of a shared understanding* of *what exactly constitutes proper evidence and rationality* in modern fundamental physics.

Some have begun to argue that physics again needs History and Philosophy of Science. For example, Nobel laureate David Gross used to think (echoing pronouncements of famed 20[th] century particle physicist Richard Feynman) that physicists need "philosophers and historians of science like birds need ornithologists"—yet, he changed his mind: "we need each other", Gross now believes (Wolchover 2015). Indeed physicists have begun to turn to the history and philosophy of science to navigate the problems signalled above. Concretely, cosmologists George Ellis and Joe Silk (2014) have argued in *Nature* that philosopher Karl Popper's criterion of testability needs to be (re)instated to demarcate proper cosmology from unwarranted theorizing, such as the construction of multiverse models. At the same time, string theorists like Gross are agreeing with philosopher Richard Dawid (2013), who claims that string theory necessitates a novel understanding of scientific rationality—an understanding in which progress can be attained on the basis of theory alone.[8]

---

[4] Hossenfelder (2017); On the history of the concept of 'fact' in physics, see Ten Hagen (2019, 2021) and De Waal and Ten Hagen (2020).
[5] Nima Arkani-Hamed, quoted in Blom and Wessel (2017).
[6] Heilbron (1992); Kragh (1999).
[7] See e.g. Kaiser (2002, 2005, 2011).
[8] David Gross for example extensively discussed and endorsed Dawid (2013) at the 2014 Strings conference in Princeton in his plenary lecture (27 June, 2014, Princeton University), in response to criticism that string theory lacks empirical connections; see https://mediacentral.princeton.edu/media/1_zmp4cbpn (consulted on 4 September 2020).

This discussion, however, has suffered from the absence of a sufficiently historiographically informed perspective. As Thomas Kuhn (1977a) already argued, and as for example the history of dark matter research illustrates, science, and cosmology in particular, did not progress via Popperian falsification tests: often, theoretical insight preceded observation, and for prolonged periods, observation may look entirely equivocal or even be absent. In the case of dark matter, existing observations were only construed as evidence for dark matter once institutional arrangements (the creation of the discipline of physical cosmology) and theoretical preferences (the wish for a closed universe) were aligned such that that interpretation became viable—only then were data on galaxies' rotation curves and the dynamics of their clusters conceived as convincing markers of the presence of dark matter. So, this is not a story of progress via Popperian falsification attempts (see de Swart et al. 2017). Kuhn (1977a) likewise argued that astrology, for example, was falsified many times over, yet remained an established empirical source of knowledge for many centuries. In Kuhn's analysis, the demise of its authority was rather due to its failure to produce a puzzle solving tradition in which theory could be tweaked or replaced. Astronomy, however, retained authority, even though plenty of its pronouncements (like those of, say, meteorology or medicine) were regularly falsified too. What these examples illustrate is that attempts at falsification are simply not the method by which, historically, science has typically progressed, and that it is not by Popperian empirical testability that scholars have 'demarcated' science from non-science in actual practice.

Furthermore, if one were to use empirical testability as a demarcation criterion of proper scientific practice, as Ellis and Silk suggested, then one would need to have a shared sense for when a hypothesis is 'testable'. Yet, in modern cosmology judgment on what theoretical hypotheses may be considered as falsifiable varies considerably. For instance, the hypotheses of dark matter and dark energy have been judged to be *conventionalist*, i.e. illicit escapes from falsifying experiments in Popper's terminology[9]—the same point that critics also level against the inflation scenario. Yet, dark matter, dark energy and inflation are all concepts that George Ellis, in a separate publication (2017, p. 24), finds "well-established scientific theory". We may conclude that a call for testability will likely fail in the concrete case of strings and modern cosmology: physicists themselves need not agree what are in fact examples of testable science. Judgment on testability of a hypothesis itself appears as theory laden and dependent on the perspective of the scholar making it.

The example of Copernicus is particularly apt for today's discussion. Copernicus proposed his alternative to the fairly successful Ptolemean universe in 1543. Yet, this theoretical proposal was basically beyond any meaningful notion of empirical falsifiability. This situation persisted pretty much until Galileo pointed the newly invented telescope to the heavens and in 1610 observed the phases of Venus. Following Ellis and Silk's prescription for modern cosmology, however, one would conclude that Copernican hypothetical theory building ought not to be identified as proper scholarship for many decades. That seems awkward: rather, this historical example teaches us, as Kuhn tried to teach Popper, that testability is insufficient as a marker of proper cosmological practice.

---

[9] See Merritt (2017).

Yet, Galileo's observations were *not* immediately awarded authority, and Copernicus' ideas were of course notoriously controversial—but not because of their non-empirical nature: Galileo still needed to *establish the authority* of empirical methods. To do so, current historiography argues, he also needed to establish the authority of a novel kind of scholar: that of a 'mathematical philosopher', who believed that valid knowledge could be expressed in mathematical laws, and attained by observation and experiment.[10] This kind of identity formation and demarcation process, we will argue in this essay, we now see signs of in the current dispute about string theory. That observation may yet offer insight into the nature of that debate, and how to appreciate it in the dynamics of scientific innovation. To do so, we will briefly touch upon Einstein and his biography—at least, on how this functions as a resource in current debates about *who is a proper theorist*, and thus, who holds epistemic authority. We will also briefly return to the example of Galileo.

But let us first digress a little more in philosophy: if Popper's falsificationalism can not act as a referee in the string theory debate, what, indeed, about Richard Dawid's approach? Have argumentative meta-structures for string theory been identified that we can warrant epistemologically? Is, then, belief in progress by theory alone justified?

**Contra Richard Dawid, Karl Popper and other Viennese**

Richard Dawid, in his important book *String theory and the scientific method* (2013), has identified three types of non-empirical argument that explain trust in string theory. These are 1) lack of viable alternatives that offer a consistent theory of all fundamental interactions, including quantum theory and gravity ('no-alternatives argument', or "NAA"); 2) the 'unexpected explanation' (or "UEA") of facts and concepts that follow from the theory, like black hole entropy; and 3) the 'meta-inductive' notion that string theory is modelled on earlier theories that did find direct empirical warrant after first being endorsed due to a lack of known satisfactory alternatives ("MIA"); these earlier theories should presumably include the Standard Model of particle physics and general relativity—yet, this seems awkward, since e.g. Einstein was live testing general relativity to the observed perihelion motion of Mercury as he progressed to his theory, while alternatives circulated. In any case, Dawid has additionally claimed that these three types of arguments deserve further confidence, as one can offer Bayesian accounts that suggest that the likelihood of a certain theoretical hypothesis increases if it is of one of the above types.[11] Finally, as pointed out already, he believes that in string theory, by virtue of the authority that its adherents award the above type of arguments, we are facing a novel kind of scientific practice: "a shift in the balance between theoretical and empirical elements can be observed," even if, when available, the empirical is still the final arbiter.[12]

---

[10] On Galileo, see Biagiloi (1993), Heilbron (2010); on the introduction of the experiment and identity of the experimenter in the early modern period, see also, e.g., Shapin (1994).

[11] Dawid et al. (2015); Dawid (2020).

[12] Dawid (2013, p. 2) argues that in theory assessment, criteria "have been significantly transformed in fundamental physics in recent decades." His 2013 book aspires to identify "the conceptual basis for the new aspects of scientific reasoning and […] why this altered conceptual basis has occurred." Indeed (p. 4), "the book's

Yet, already the example of Copernicus suggests that the sentiment that string physicists' weighing of theoretical evidence is without precedent may not hold up to the scrutiny of historical comparison. This, I imagine, could be a reassuring observation for string theorists—yet, it also deserves to be pointed out again that empiricism had little authority at the time of Copernicus. Furthermore, it is generally wise to be mindful of the inevitably particular and contingent character of each and any individual historical case, before drawing general conclusions.

In any case, it would be a moot point whether the example of Copernicus would or would not fit Dawid's NAA, UEA or MIA, and Dawid's analysis does not aim at historiography. Rather, his arguments are focussed on theory assessment: Dawid has primarily aimed to identify and offer epistemic license for certain kinds of non-empirical reasoning. Yet, his arguments have been drawn into debates about demarcation and justification of research programs, as they are held to assist in deciding on a course of action when doing research: persist in string theory and multiverse modelling, or not.[13] It is for this reason that they have attracted the attention of Gross and string critics like Ellis and Silk.

Such an application, however, reminds us of the old separation between contexts of discovery and context of justification, which presupposed that one can assess a theory's viability independent of the contingencies of the work done to construct it. The distinction was once introduced by Hans Reichenbach and Karl Popper to separate for philosophy the project of justification, that would be done objectively, from the subjective and logically opaque nature of actual research.[14]

The suggestion of objectivity will make it attractive to string theorists such as Gross to invoke Dawid's arguments to justify string theory's dominance. Yet, a neat separation between the context of discovery and the context of justification, Thomas Kuhn complained, "does not fit observations of scientific life."[15] As he elaborated after comparing historical cases of theory choice, such as phlogiston versus Lavoisier's oxygen or Copernicus' versus Ptolemeus' solar system: "Considerations relevant to the context of discovery are […] relevant to justification as well; scientists who share the concern and sensibilities of the individual who discovers a new theory are ipso facto likely to appear disproportionally frequently among that theory's supporters"—just like David Gross in the case of string theory. Kuhn's examples were meant to illustrate that a choice of theory often only appears clear in hindsight: 'crucial' experiments are only performed or recognized as such once a theory has already become established. Clean logical presentations for the cases of Copernicus or Lavoisier typically ignore the incommensurability of competing positions and the messy situation of prolonged stasis and stalemate in which evidence is entirely equivocal. Kuhn: "That is why it has been

---

core message" is that "the novelty of current theories in fundamental physics is not confined to the conceptual level of those theories themselves. Rather, it extends to the meta-level of theory assessment where a shift of the balance between empirical and theoretical can be observed." On the point of the novelty of string theory methodology, see also Dardashti and Hartmann (2019), in particular pp. 3-5.

[13] See Dawid (2019) for his discussion of this point.
[14] For a recent discussion, see e.g. Arabatzis (2006).
[15] Kuhn (1977b), p. 327.

difficult to construct algorithms for theory choice." [16] History taught him that choice of theory does not follow the neat recipes that formal epistemologists, Bayesians included, were trying to pin down.[17]

Why bring up Kuhn's criticism of Popper and the logical empiricists of more than half a century ago in today's discussion of string theory? Because the positions of Ellis, Silk and Dawid echo those of the earlier period. Popper and Reichenbach wished to resolve the problem of induction and the challenges that it produced for empiricist accounts of demarcation and confirmation: Reichenbach proposed a probabilistic and Bayesian perspective on the latter, while Popper tried to circumvent the issue by emphasizing falsification instead of confirmation in methodological assessment. Both were addressed by Kuhn's *Structure of Scientific Revolutions* (1962), which implied that these perspectives fail in comparison with the actual historical record: they do not capture how choice of theory is made in the actual practice of science. In Kuhn's analysis, science is conducted in paradigms that colour and condition both experiment and assessment, as illustrated by his famous duck-rabbit analogy. Theory choice, Kuhn's historical examples of Darwin, ether and phlogiston aimed to exemplify, takes place in a way that resembles the weighing of values in moral judgment, rather than that it follows falsificationist prescriptions or Bayesian-style recipes: one weighs certain theoretical virtues, just like one weighs moral values.[18] Even if today's historical study of science has moved beyond Kuhn when it emphasizes the conditioning done by scientific practice over the role played by paradigm, it has moved in the opposite direction of Popper and Reichenbach.

So, did Dawid succeed where, if one were to follow Kuhn, Karl Popper, Hans Reichenbach and other Vienna-style twentieth century empiricists failed? That is: Did Dawid (who, true to form, also hails from Vienna) succeed in offering an account of theory choice for the case of *non*-empirical science, while his predecessors still aimed to capture and validate a version of the empirical method?

Given Kuhn's criticism of such accounts in the case of empirical science, it may be wise to lower expectations when they are applied in the case of non-empirical science and string theory. I do not wish to criticize Dawid by lobbing up the familiar arguments against Bayesianism (as there are: prior probabilities, etc.[19]), and I am happy to acknowledge that UEA, NAA and MIA do capture some elements of the argumentative structures that play a role in assessing string theory. Yet, I also believe that historiography teaches that we cannot expect such reductionist accounts to capture the full *practice* of theory choice in physics—or, to put it differently, they do not represent some boiled down essence of, in Kuhnian parlance, the full dynamics of paradigms. Since the time of the logical empiricists, historiography of actual science has taught that that broader kind of a perspective is needed to understand how scientific rationality does function. Dawid's arguments may thus also be of limited value in understanding what is at stake in debates about post-empirical physics and the status of string theory as proper science.

---

[16] Kuhn (1977b), p. 328.
[17] On the insufficiency of Bayesian accounts of theory choice, see Kuhn (1977b), pp. 328-329; for an attempt to bridge the differences between Kuhn and logical empiricism via Bayesianism, and an indication of its inherent limitations, see Salmon (1990).
[18] See on this point particularly Kuhn (1977b).
[19] See e.g. Talbott (2008) for an overview of the subject.

A relevant point to observe is that there is hardly a theorist today that does a Bayesian calculation before deciding on a theory or research program—and there is no theorist who finds that kind of analysis convincing when it goes against his or her existing convictions and commitments, just as Kuhn argued in the case of more straightforwardly empirical science. This is illustrated by how two prominent adherents of a competitor program to string theory, Carlo Rovelli and Lee Smolin, have taken exception to Dawid's analysis itself. They deny the applicability and authority of NAA and MIA in favor of string theory, as they claim that the same arguments work just as well for their loop theory of quantum gravity, and that, in any case, Bayesian confirmation and non-empirical arguments are epistemically far too weak to license belief.[20] Their positions may well be countered[21]—yet, from Kuhn's perspective, such objections are exactly the kind of point you expect scholars in a competing paradigm to bring forward.

Then, the above may suggest that, rather than formal epistemology, a perspective that addresses theory choice as rational but embedded in the dynamics of paradigms and the conditioning of practice may be better suited to offer understanding of the debate on non-empirical physics. One instrument to study practice in relation to theory assessment is the perspective of *epistemic virtue*. This focuses on the virtues (broadly, not just morally understood) deemed necessary for a scholar to attain knowledge. As such, it reflects the culture of doing science related to a particular paradigm.

A focus on virtue has received novel impetus in history and philosophy of science,[22] as may be illustrated by the recent contribution by D.J. Hicks and T.A. Stapleford. They have argued, drawing on work by moral philosopher Alisdar MacIntyre, that to understand the communal and normative nature of scientific practices, we should appreciate that certain dispositions to think, act and feel according to particular models of excellence are expected: forming a scientific practice involves broad ethical stances and accounts of virtue. When these are recognized, there will be confidence that particular scholars aim at the intrinsic epistemic values of a discipline; and that they weigh these appropriately, for instance in relation to personal rewards. Such dispositions are taught, and they are debated, particularly when a discipline's boundaries are at stake. In the case of controversies, debates about the future of a practice often entail competing historical narratives. As Hicks and Stapleford put it: "arguments over the present always entail reinterpretations of the past."[23]

Examples are easy to find: in the late nineteenth century, British debates on the proper role of the *imagination*, versus observation and experiment, manifested themselves as, e.g., discussions about who was to be assigned priority for the law of energy conservation—Julius

---

[20] Cited on p. 2 and p. 1 in Dawid (2020); Smolin (2014); Rovelli (2019).
[21] Dawid (2020, pp. 2-5) discusses Smolin's criticism.
[22] For references that particularly relate to historiography, see for example Daston (1995); Söderqvist (1997); Stump (2007); Daston and Galison (2007); Kidd (2014); Bellon (2015); Stanley (2017); Pennock (2019); for an overview of primarily philosophical literature, see Turri et al. (2017).
[23] Hicks and Stapleford (2016), p. 472.

Mayer or James Joule. In a related development, energy physicist P.G. Tait vehemently debated social philosopher Herbert Spencer on the merits of a priori reasoning; they particularly contested aspects of Isaac Newton's biography, and argued over the correct reconstruction of Newton's priority debate with Leibniz. In these debates, attributions of virtue and vice were essential to demarcate proper practice from false avenues.[24] As we will see, varying accounts of Einstein play a similar role in today's dispute over the empirical.

Of course, historiography with a focus on virtue is hardly a panacea to answer any or all questions about proper practice for today's physics—yet, as the above suggests, it may give us helpful insight into the process of identity formation and the establishment of authority in scientific innovation. These aspects do play a role in the current assessment of string theory—that is, in the assessment of string theorists *as physicists*.

**Authority and Identity**

Is string theory proper physics? In history of science, one has generally shied away from taking a normative role, given the contingent character of past science: History does not prescribe solutions, even if one does learn from history.[25] One way to contextualize the question above, and to make it more amenable to historical analysis, is to translate it into the question: are string theorists *proper physicists*? That question reflects indeed one level at which the debate is taking place: debates about demarcation are inevitably debates about identity, in relation to authority. These questions are also tied to discussions about funding, which helps to explain why the issue is so contentious.

One way to address the above questions is to address scientists' beliefs regarding the correct *way of being* to attain knowledge—how one believes knowledge is attained and what norms and dispositions this requires: what *epistemic virtues* one subscribes to and what role they play in *doing* theoretical gravity studies. Ideals of epistemic virtue are intimately tied up with the practices performed and theoretical principles held dear: both are associated with one's sense of professional identity. They are reflected in the idealized yet abstract concept of the '*persona*': the archetypical 'scientist', 'physicist' or 'theorist' that is internalized and adhered to as an example. Thus, they have an immediate effect on the production of knowledge.[26] As we will see, these may also inform and reflect attitudes towards the empirical.

What epistemic virtues are observed by the relevant scholarly groups, and how have they grown historically? What virtues and norms have theoretical physicists perceived to hold *scientific authority* before—what were the norms and values by which they have held in the past that knowledge can be attained? How has that translated into the weighing of various kinds of principles, evidence and the instantiation of different knowledge practices? To address these questions, one needs to see what the ideal has been of a *good physicist*, or, indeed, a *good scientist* and *how that ideal has evolved*.[27]

---

[24] Yeo (1988); Smith (1998); Higgitt (2007); Saarloos (2017); Saarloos (2021).
[25] See for instance Kragh (2019), who offers the example of Victorian soliton-like atom building as a forewarning for string theory attempts; for a general discussion of how historiography may inform the present, see Guldi and Armitage (2014); for a discussion of its relevance to history of science, see the discussion section in *Isis* 107 (2), pp. 309-357, 2016.
[26] On the role of epistemic virtues and persona in scholarship, see e.g. Daston & Sibum (2003); Daston & Galison (2007); Shapin (2008); Paul (2011); Wang (2012); van Dongen and Paul (2017); Tai (2017).

These questions can inform us in the debate about the empirical. For instance, today string theorist Leonard Susskind and cosmologists Ellis and Silk will agree that the persona of the 'philosopher' is to be shunned by proper physicists. Yet, Ellis (2017) expresses that string theory cosmology is to be "regarded as philosophy", and not science, while Susskind, invoking the authority of his professional identity as a physicist, dismisses demands for empirical falsifiability as merely 'philosophical' (Kragh 2014a). Already in 2005, he dismissed such demands as naïve and denounced those who raised them as "Popperazi" (Gefter 2005). To Susskind 'confirmation' is what really matters, regardless whether that is exclusively theoretical. Indeed, string theorists find confirmation when observing agreement amongst themselves regarding a particular hypothesis (Polchinksi 2019a). Their critics, however, point out that such a culture of conformity leads to group-think (Smolin 2006a).

Dissent does not fare well in today's funding arenas. It is not hard to imagine that many referees like to reward proposals in which they see their own opinions reflected;[28] a tendency to confirm rather than to disagree is then effectively a beneficial trait, even if 'conformity' will generally be considered a vice. This circumstance may amplify certain practices of rationality or particular forms of knowledge instead of others: conformity favors further calculation of established ideas over disputes concerning principle.[29] Stances on proper epistemic attitudes—that are related to ideas about what are ideal scholars, deserving of funding—clearly matter for the kind of rationality and evidence considered. This is a consequence of their intrinsically social nature.[30]

Ideals and norms are increasingly *globally shared* as science itself has moved from the *local* to the *global* in the last half century. It has done so particularly in the field of theoretical high-energy physics, which was the first to introduce global online preprint publishing in 1991: the 'arXiv'. This accelerated an already strong tradition of international collaboration and refereeing, in part promoted by Cold war contexts and the physical scale of high energy experiments.[31] The above developments, in the context of centralized funding competitions, do suggest the existence of institutional conditions amenable to the *harmonization of epistemological stances*,[32] and the marginalization of dissenting points of view and subcultures: they promote situations in which, indeed, there is 'no alternative'. Such a dynamic, however, remains hidden from view in formal confirmation theory. One may then hesitate to follow Dawid, who takes the point that a large majority believes that alternatives are lacking as a starting observation for the Bayesian warrant offered by his No Alternatives Argument for string theory.

---

[27] We strongly sympathize with the approach taken in the study by Gilbert and Loveridge (2020), who discuss differences in epistemic, social and physical positions between string theorists and loop quantum gravity theorists. Their synchronic-statistical (rather than diachronic and historiographical, as argued here) survey focuses on differences in physical, social and epistemic "tastes". It shows that historical conditioning (with string theorists inheriting positions from particle physics, and loop theorists inheriting positions from relativity) are highly determinative of correlative positions.

[28] For recent modeling on harmonization of epistemological stances due to competitive funding practices, and further references on this point, see Avin (2019).

[29] Gilbert and Loveridge (2020) argue that their statistical survey of quantum gravity scholars reveals "a confirmation of objectivity through intersubjectivity"; objectivity "manifests itself as standards for stylistic consilience." They found that "this mode of investigation existed predominantly, though not exclusively, in string theory as compared to loop quantum gravity" (pp. 23-24).

[30] Likewise, Bezuidenhout et al. (2019) argue that an imbalanced role of the virtue (or vice) of 'docility' in university education has lead to an unwarranted and uncritical dependency on established experimental protocols in the life and social sciences, and point out that this plays an important role in the current "replication crisis" (p. 82). They further insist that virtues need to be appreciated in their social embedding, for instance in the teaching context.

Scientific innovation can be accompanied by the introduction of *novel ways of being*. These also introduce conflict of authority and identity. In this regard, the string debate mirrors controversies surrounding Einstein and Galileo, who themselves were part of a process of forging new identities, precisely along the theory vs. experiment dichotomy: Einstein was both the icon and iconoclast example of the novel theoretical physicist, while Galileo established the authority of the 'mathematical' philosopher who trusted experiment and observation. Their innovations, too, were accompanied by questions about the authority of their novel brands of scholarship; in their cases, just as in the example of string theory, debates raged whether their kind of knowing was *proper*: whether Einstein was really a proper physicist, and whether Galileo should be awarded the authority of the philosopher. We will return to the case of Einstein in the next section, but it is instructive to first briefly pay some more attention to the example of Galileo.

Galileo started off as a professor of mathematics in Pisa, which put him pretty low on the social-epistemological ladder of his time: somewhere among the musicologists and engineers, and definitely a few rungs below proper philosophers. He debated those on a number of occasions, for instance on the subject of buoyancy, when he reproached the Aristotelians for their ignorance of mathematical methods and general lack of novelty. Yet, both were foreign to the socio-professional ethos of the Aristotelians; embracing these would mean to award them an authority they lacked in their world, thus unduly elevating Galileo's kind of scholarship. The debate had moral overtones—slavish Artistotelianism was contrasted with narcissistic daring—that signal the demarcation of identities, and, in the words of historian Mario Biagioli, "the possible emergence of incommensurability."[33] In the end, by becoming *court philosopher* at the Medici court in Florence, Galileo fashioned a position that came with the power and opportunity to gain acceptance and authority for a novel kind of epistemic identity: that of a scholar who trusted mathematical and empirical methods first, and who argued his positions with a heavy dose of wit and rhetorical sharpness. The novelty and controversies associated with empiricism were appreciated at court, but not at traditional universities. To secure the position, Galileo had prominently introduced some of his newly discovered celestial bodies into the local Medici mythology, thus boosting the social standing of his observational methods—yet their authority was not absolute and was still highly precarious, just as his own, as he subsequently found out in his run in with the church and its scholars.

---

[31] On 20th century globalization of science in the cold war context and the role of physics in it, see e.g. Krige (2006); van Dongen (2015); Krige (2016).
[32] Avin (2019).
[33] Biagioli (1993), p. 236.

In today's debate, Ellis and Silk (2014, p. 323) have made the point that "journal editors and publishers could assign speculative work [such as string theory and multiverse cosmology] to other research categories—such as mathematical rather than physical cosmology—according to its potential testability." Thus, string theorists and multiverse cosmologists would be denied authority as physicists. Such arguments do suggest that, as in the case of Galileo, a new identity is indeed forming, centering on 'non-empirical' physics— even if, at the moment, despite its successes and institutional dominance, this finds itself in "a no-man's land between mathematics, physics and philosophy that does not truly meet the requirements of any" in the judgment of Ellis and Silk (p. 321).[34] The subject still needs to find its social space and the associated epistemic authority.

So, paradigm changes can be accompanied by the creation of novel scholarly identities, and a related debate over values and virtues; this helps to explain why Kuhn argued that choice of theory is *like* moral judgment. A focus on epistemic virtue makes that link even more direct, as the familiar example of objectivity illustrates: the virtue of objectivity implies a moral economy of the self, just as much as it constrains scientific practice.[35] It is not hard to imagine a similar functioning for particularly theoretical virtues—'simplicity', e.g. in relation to 'unification'—and indeed they have functioned as such, for example in the case of Einstein. His example is in fact particularly relevant, as Einstein also debated the merit of a version of non-empirical physics. Furthermore, he often features in current discussions on string theory, invoked by both its proponents and their critics.

**Einstein and non-empirical unification physics: epistemic virtues**

Today, among scientists as much as the general public, the quintessential example of a *physicist* is of course Albert Einstein. Among physicists themselves, he is furthermore the leading example of a *theorist*, particularly due to the successes of his theory of relativity. Yet, both the authority of the theory of relativity, and Einstein as a scholar, were debated at various times early in the twentieth century. Debates ranged from mild mannered exchanges with e.g. Hendrik Antoon Lorentz on the merits of the ether,[36] to tense standoffs with reactionary 'anti-relativists' like experimentalist Philipp Lenard. Let us first briefly consider this second episode.

In the late nineteenth century, the subdiscipline of theoretical physics was a fairly recent phenomenon. It had come into existence following the creation of novel professorial chairs that were meant to alleviate the teaching load of professor-directors of university research labs.[37] Originally in a subordinate position, the subdiscipline grew in esteem and authority, particularly due to the development of the electron theory, quantum physics and relativity—which accelerated following the public success of Arthur Eddington's eclipse observations of 1919.[38] This produced a backlash in the next decade and beyond: noted experimentalists as Lenard and Ernst Gehrcke took exception, and, joined by conservative philosophers and engineers, an anti-relativity movement ensued. Einstein represented much that was disliked about the post-WWI order: as a leftist and Jew, he quickly became identified as yet another cultural revolutionary, who in his case sought to bring down long-established concepts as the ether and preferred frames of reference. Einstein did so with a particularly *mathematical* theory, his critics pointed out. That was one reason why they were unwilling to award him and his work epistemic authority: Einstein's abstract contribution was unnatural and superficial. Its rise to prominence was an expression of a moral decay in society, which, with the ascent of theoretical physics, prominent in liberal Berlin, and the perceived marginalization of experiment and experimenters, was seen to have expanded into physics.[39] This example once again illustrates that in paradigm change, judgments of theory, and of what it means to be a proper scientist, stand in an inextricable relation to one another. In the end, of course, it was exactly Einstein's relativity that established him as a novel disciplinary icon.

---

[34] For a discussion of string theory as wedged between physics and mathematics, see Galison (2004); an earlier study of its relation to the empirical is Galison (1995); for more on the string theory debates and boundary discourse, see Ritson and Camilleri (2013).

[35] Daston and Galison (2007); a similar argument can be made for the empiricist virtue of *exactitude* that was taught across the sciences and humanities in 19th century Germany; see ten Hagen (2021).

[36] See e.g. letters contained in Kox (2008) and the discussion in Kox (1988; 2019, pp. 68-73).

Let us leave the case of the anti-relativists. There is another instance in Einstein's biography in which the identity of the theorist was an issue, and in that case, he himself articulated how *non-empirical* theory construction was a valid pursuit; he did so by arguing for the epistemic benefit of its practice—which, in his judgment, equally demanded a particularly virtuous disposition. The case at hand is Einstein's search for a *unified field theory* that he undertook for many years.[40] Its goal was to unify in one mathematical description the classical gravitational and electromagnetic forces, in such a way that the observed quantum nature of matter was reproduced, particularly through the derivation of some solitonic elementary particle solution of a novel set of field equations. The effort was usually conducted without any concrete empirical input, and did not lead to observable predictions; it was primarily mathematical in nature. Earlier in his career, as in the examples of his light quantum hypothesis or the electrodynamics of moving bodies, Einstein's theory construction did directly engage with the empirical. He himself observed a shift, too:

> Coming from a skeptical empiricism à la Mach, I have become, through the problem of gravitation, a converted rationalist, that is, someone who believes that the only reliable source of truth can be found in mathematical simplicity.[41]

---

[37] Jungnickel and McCormmach (1986).
[38] On this point, see e.g. Staley (2008); Beyerchen (1977).
[39] The literature on anti-relativists is substantial, see e.g.: Forman (1984); Schönbeck (2000); Rowe (2012); van Dongen (2007; 2012); Wazeck (2013).
[40] The following is based on van Dongen (2017); for a more straightforwardly historical-technical treatment of Einstein's work on unified field theory without the emphasis on virtue, see van Dongen (2010). Other pertinent discussions of Einstein's unified field theory work are found in Vizgin (1994); Sauer (2006, 2014); Goenner (2004; 2014).
[41] Einstein to C. Lanczos, 24 January 1938, Einstein Archive, Jerusalem, call number 15-268 (cited on p. 259 in Holton 1988).

Einstein's construction of the general theory of relativity used ideas and methods that one can identify as typically grounded in physics (e.g. the equivalence of gravitational and inertial mass, Newtonian gravity), and ideas and methods that were more directly derived from a mathematical body of knowledge (general covariance, the Riemann tensor).[42] In his later recollections of his road to general relativity, made when he pursued unified field theory, Einstein over-emphasized the role of mathematics, at the expense of the contributions that followed from his physical strand.[43] Striving for 'mathematical naturalness' and 'mathematical' or 'logical simplicity' were terms that he used more or less interchangeably when discussing his unification attempts and his recollections of formulating general relativity. For example, Einstein wrote to David Bohm in 1954:

> I believe that these [unified field theory] laws are logically simple and that the faith in this logical simplicity is our best guide, in the sense that it suffices to start from relatively little empirical knowledge. If nature is not arranged in a way corresponding to this belief, then there is no hope at all to arrive at deeper understanding. […] This is not an attempt to convince you of anything. I just wanted to show you how I came to my position. The realization that in a half empirical way one could never have arrived at the gravitational equations for empty space had a particularly strong influence on me. In that case, only the viewpoint of logical simplicity can be of help.[44]

But Einstein did want to convince Bohm of something: namely that his efforts in unified field theory, controversial among the latter's generation of theorists,[45] were a valid attempt, despite its record of failure and problematic relation to the empirical.[46]

To drive his point home, Einstein enlisted his own biography and in particular his success with general relativity, aware of his stature among Bohm's generation.[47] He did so in a way that emphasized and tied together the epistemological benefit and required personal virtue of these efforts. For example, Einstein argued that trusting the empirical—as quantum physicists had done—while believing that "facts by themselves can and should yield scientific knowledge" was grounded in a "prejudice."[48] To change theories because of a conflict with experiment was "trivial, imposed from without"; striving after unification and logical simplicity followed a "more subtle motive."[49] Similarly, Einstein confided to the young mathematical physicist André Lichnerowicz that if the unification program would fail, understanding "the physical world [...] superficially" was all that remained, just as he wrote to Max Born that mere "humans are usually deaf to the strongest arguments, while they are constantly inclined to overestimate the accuracy of measurement."[50] Yet, "a true theorist" is convinced that comprehension is "built on premises of great simplicity" (Einstein 1950, p. 13): unifying is to aim for *true* understanding, and not the superficial of the empirical; to unify required to stand firm against the vices of prejudice and superficiality.

---

[42] See Janssen et al. (2007).
[43] See e.g. Einstein (1933); van Dongen (2010). John Norton (2000), however, does not believe that Einstein's later recollections are off; yet Janssen and Renn (2007) do; Renn and Sauer (2007) offer a treatment of Einstein's road to relativity that is less pronounced on this issue.
[44] Einstein to D. Bohm, 24 November 1954, Einstein Archive, Jerusalem, call no. 08-054 (cited on pp. 181-182 in van Dongen 2010).
[45] For critical attitudes among Bohm's generation of Einstein's unification attempts, see e.g. the judgment of Abraham Pais (1982), pp. 325-354; pp. 460-469, in particular p. 325 and pp. 349-350; and Robert Oppenheimer, cited in: Bird and Sherwin (2006), p. 64; Schweber (2008), pp. 276, 279.
[46] Einstein (1933) argued in his lecture at Oxford University on methodology of theoretical physics and his unification attempts that the empirical was primarily involved in testing of theory, but that the road from the empirical to the formulation of theory was rather tenuous and insecure (see also Norton 2000; van Dongen 2004). He did, however, during the 1920s hypothesize the existence of 'ghost' charges and other possible material-inertial explanations for the geomagnetic effect, and considered to accommodate for those in field equations of a unified field theory. In this case, the empirical then did contribute to the formulation of theory in a more direct fashion. The attempt did not bear any fruit; see Illy (2019).
[47] To get a sense for how high Einstein was held in esteem by the next generation of physicists, see e.g. Heisenberg (1989).
[48] Einstein 1949, p. 49.

So, Einstein's epistemic convictions were intimately related to his sense of personal virtue. Striving for understanding itself was virtuous, as Einstein wrote in an essay on '*Good and evil*' (1934): "[I]t is *not the fruits* of scientific research that elevate a man and enrich his nature, but the *urge to understand*, the intellectual work." Furthermore, finding understanding in physics implied unifying: Einstein, in his own words, concerned himself "first and foremost" with the "logical unity in physics [...] to understand the existing, real world. [...] [T]he totality of logically independent axioms, stands for the 'un-understood remainder'."[51]

His unified field theory efforts may not have been successful, but as they aimed at true understanding, they offered elevation—and, possibly, personal deliverance and liberation—as one tried to "arrive at those universal elementary laws from which the cosmos can be built up by pure deduction."[52] In an article entitled '*Religion and science*', he expressed that "individual existence is a sort of prison and one wants to experience the universe as a single significant whole"; aspiring to grasp this "cosmic religious feeling" was "the strongest and noblest motive for scientific research. Only those who realize the immense efforts and, above all, the devotion without which pioneer work in theoretical science cannot be achieved are able to grasp the strength of the emotion out of which alone such work, remote as it is from the immediate realities of life, can issue" (Einstein 1930, pp. 41-42).

The personal virtue and promise of deliverance that Einstein ascribed to attempts to find a unified description of the universe resonated with his Spinozism: all who had made similar attempts and achieved any success, had equally been "imbued with the truly religious conviction that this universe of ours is something perfect and susceptible to the striving for knowledge. If this conviction had not been a strongly emotional one and if those searching for knowledge had not been inspired by Spinoza's *Amor Intellectualis Dei*, they would hardly have been capable of that untiring devotion which alone enables man to attain his greatest achievements."[53] Spinoza, in his *Ethics*, had argued that man's moral well-being lies in a life spent on the pursuit of knowledge, particularly knowledge of the eternal laws of nature—that is, the laws of an infinite yet deterministic universe, or, synonymously for Spinoza, God: the better we understand His true ideas, the more equipped we are in dealing with life's disturbances. Einstein famously agreed with Spinoza's view of an impersonal yet existing God, and shared with him a sense that religiosity was really the "confidence in the rational nature of reality as it is accessible to human reason. Wherever this feeling is absent, science degenerates into uninspired empiricism."

---

[49] Einstein 1950, p. 16, p. 13.
[50] Einstein to A. Lichnerowicz, January 1954, Einstein Archive, Jerusalem, call no. 16-319; cited in van Dongen (2010), on p. 5; Einstein to Max Born, 12 May 1952, cited in van Dongen (2017), p. 68.
[51] Unpublished manuscript by Einstein, dated 1931, Einstein Archive, Jerusalem, call no. 2-110.
[52] Einstein (1918), p. 247.
[53] Einstein (1948), pp. 56-57.

Following the methodology of mathematical unification was also epistemically successful, Einstein claimed. For example, in his Spencer lecture at Oxford university in 1933 'On the method of theoretical physics', Einstein argued that the "simplest", i.e. unified, equations for 'semivectors' (these were his alternatives for spinors) "furnish a key to the understanding of two sorts of elementary particles, of different ponderable mass and equal but opposite electrical charge [i.e. electrons and protons]."[54] The result proved spurious,[55] however, as was usually the case in his unified field theory efforts. Most importantly, he would regularly make the methodological point by offering his recollection of his path to general relativity—as he wrote to his collaborator on semivector theory, the mathematician Walther Mayer: "one should look for the mathematically most natural structures, without initially being bothered too much about the physical, as this brought the desired result in gravitation theory."[56] Thus, he enlisted his biography, and his iconic and exemplar status among theorists, to argue for his kind of unification physics. He offered himself, his version of how general relativity had been found, as an ultimate and indeed virtuous example—as exemplary authority—in his disputes on the proper nature of how to do theoretical physics. These disputes addressed whether unified field theory efforts had a proper place in theory: they did, so the argument would go, because he was a proper theorist, of proven sound judgment and appropriate dispositions.

Yet, many of his peers disagreed, particularly on the latter point—when it came to Einstein's unification physics. The younger generation of quantum physicists—just like the philosophers of the Vienna circle—had particularly admired Einstein as the empiricist icon that had created the special theory of relativity, in which the coordination of spacetime points to events was given by observational operations.[57] The general theory was of course considered Einstein's greatest triumph, and his recollections of that period were not disputed, but his being an exemplary unification physicist, remote from the empirical, was. Dismissal could take a moral tone, for instance when Robert Oppenheimer deemed that Einstein had been "wasting his time." In fact, he had gone "completely cuckoo", Oppenheimer added in private, or, as he put it in public, Einstein had "lost contact with the profession of physics." Clearly, the Einstein of unified field theory was *not* a proper theorist.[58]

---

[54] Einstein (1933), p. 301.
[55] Van Dongen (2004; 2010), pp. 96-129 (chapter 5).
[56] Einstein to W. Mayer, 23 February 1933, Einstein Archive Jerusalem, call no. 18-163, cited on p. 119 in van Dongen (2010).
[57] For admiration of Einstein as empiricist icon, see e.g. Heisenberg (1989); Heisenberg here further recalls his surprise when Einstein explained to him in 1926 that he no longer held empiricist views. In 1927, Heisenberg signaled a difference of opinion regarding the role of 'simplicity' and the empirical with Einstein (Heisenberg to Einstein, 10 June 1927, cited on p. 467 in Pais 1982); Einstein himself was well aware of his isolation and the negative judgment of his peers; see Pais (1982), p. 462. See Howard (1994) on the logical empiricists.
[58] Bird and Sherwin (2006), p. 64; Schweber (2008), pp. 276, 279; see also Pais (1982), pp. 325-354, 462.

In the absence of the empirical, Einstein emphasized the merit of his personal epistemological conviction, along with its success as documented in his version of his biography: the epistemic benefit of doing unified field theory was bound up with the virtuous dispositions of his kind of theorist. Critics like Oppenheimer disagreed: a unified field theorist à la Einstein was not a proper physicist. Likewise, string theorists believe that progress in physics is possible without the empirical, and, as we will see in the following section, that some of their peers are in fact exemplary physicists; their critics disagree, and are of the opinion that a scholar who practices science without involvement of the empirical may be many things, but not a *physicist*.

**Crisis in physics and epistemic virtue: the string theory debate**

String theorist Brian Greene (1999, p.15), in his book on "The Elegant Universe", considers Einstein's unified field theory attempts a "quixotic quest" that had indeed "isolated" him.[59] Yet, unlike Oppenheimer, Greene finds that "Einstein was simply ahead of his time", and he identifies Einstein's project with that of string theorists: "his dream of a unified theory has become the grail of modern physics." Clearly, Einstein is presented by Greene as both example and justification: the pursuit of string theory is appropriate and honorable, in light of the Einsteinian pedigree of the project. This, as we will see, is repeated by others: Einstein is invoked to argue in favor of (or indeed also against) string physics. Thus, the dispute about the authority of string theory manifests itself as a dispute about what *is* a physicist, and what the attributes of good physicists are—even if these latter are not being sourced in steep Spinozist accounts.

Conceptions of our self and of our professional identity—'who you are' and how you expect to do science—are tied to what epistemic choices are awarded authority. Consequently, our practice of rationality is reflected in the epistemic virtues we appreciate. Indeed, according to the prominent string theorist Joseph Polchinksi (2019a, 2019b; Polchinski passed away in 2018), string theorists are virtuous: they *overcome conservatism*, aim to be "true to the science"; they are "broad" and "original" physicists. Field theorist Steven Weinberg was an important example to Polchinski: Weinberg would have the *independence*, "clarity of mind" and "integrity" not to take shortcuts: when trying to find answers to theoretical problems, he would aspire to make a minimum of assumptions. String theory's critics, Polchinski found, need to be more *patient* and *open minded*:[60] when physics moved from the optical to the atomic scale, which took 300 years to achieve, it only had to cross four orders of magnitude on the energy scale; now, string theory is attempting to bridge another twenty-five orders of magnitude. Demanding the availability of "direct observation" in such a leap reflects a scientific methodology that is "too rigid".

---

[59] Greene (1999), p. 15; in Greene's opinion, the isolation was particularly due to a lack of consideration of nuclear forces. Einstein considered these to be of only secondary importance; solving the problem of unification came first.

[60] The survey of Gilbert and Loveridge (2020) taught them that string theory is (in contrast to loop quantum gravity) "more inclusive of diverse tastes and is therefore more successful in germinating research" […]. This helps to explain its epistemic, financial and institutional dominance, in their view, as string theory's "aesthetic regime is better aligned with the situational demands and compromises generally faced by late twentieth/early twenty-first century research fields". Inclusivity, however, only applies to choice of problems and methods internal to string theory, but not beyond (p. 22).

Although Weinberg's professional drive may inspire string theorists, it reminds Lee Smolin (2006a) of a counterproductive competitive streak. Smolin is less concerned about the non-empirical nature of string theory but is frustrated by its dominance. He believes that "new values and attitudes" are needed for the problems that quantum gravity poses. Smolin explicitly identifies the crisis in physics as tied to a choice of identity and virtue: according to him, physics needs a "seer", willing to question established principle; seers are "critical" but "not competitive". In Smolin's judgment, however, funding structures reward "craftspeople", who confirm the consensus with "technically sweet solutions". Dissent requires "character" and "courage", but these, Smolin believes, are under-valued.[61]

Polchinski gave another example, besides Weinberg, that inspired his efforts: Albert Einstein, and in particular the Einstein "that […] finished the general theory of relativity" and contributed to quantum mechanics—yet not so much his "philosophical bent."[62] Smolin, too, chooses Einstein as role model, but indeed rather emphasizes his "more reflective, risky and philosophical style." That would be needed today, since the "harsh and aggressive" way of working, taught at Harvard by the likes of Weinberg, has not had any successes after establishing the Standard model. The "Einsteins […] think for themselves and ignore the established ideas of powerful senior scientists."[63]

Einstein himself actually categorized scholars in fundamental physics in a way that is similar to Smolin's. In a letter to his close friend Paul Ehrenfest—both were frustrated by rapid developments towards the quantum theory—he found that:

> There are those that vex principles ["Prinzipienfuchser"] and virtuosi. All three of us [i.e. Einstein, Ehrenfest and Niels Bohr, whom Ehrenfest was arranging to meet up with Einstein] belong to the first kind (at least the two of us) and have little virtuosic talent. So effect at encounter with distinct virtuosi (Born or Debye): discouragement. By the way, it works similarly the other way around.[64]

In the same letter, Einstein expressed that he was no longer thinking about experiments on the wave and particle properties of light, and that one "will never arrive at a sensible theory in an inductive manner", even if "fundamental experiments" could still be of value—once again deprecating the quantum program's empirical slant. He identified some of its proponents (but, notably, not Bohr) as working in a way that was different from his own: Max Born and Peter Debye (a former student of Arnold Sommerfeld, like many other prominent quantum scholars) were seen as exponents of a technically savvy yet less fundamentally oriented kind of physics.[65] As it turned out, with the relocation of the epicenter of physics from Europe to the United States and the practice and scale of physics in the context of both the Second World War and Cold War, philosophical reflection on principle came to be seen as an unprofessional activity for a career-minded physicist.[66]

---

[61] According to Gilbert and Loveridge's survey (2020), loop theorists identify with "visionary" (truth oriented, monistic, qualitative) epistemic "tastes", whereas string theorists correlate strongly with "pragmatic" (i.e. quantitative, aimed at rapid discovery, pluralistic, utility focused) epistemic tastes. String theorists further show a preference for an "interdependent" rather than an "autonomous" social taste in their research.
[62] Polchinski (2019a), pp. 349, 342 respectively.
[63] Smolin (2006a), p. 294.
[64] Einstein to Paul Ehrenfest, 18 September 1925, Einstein Archive 10 111; translation as in van Dongen (2010), pp. 162-163, which also discusses the letter in the context of Einstein's attitude in unification physics.
[65] For a discussion of the Sommerfeld school as a training ground for 'virtuosi', see Seth (2010).

Smolin recognizes that trend, too, and laments it. He remembers being taught in the 1970s at Harvard, where "the spirit was pragmatic; 'Shut up and calculate' was the mantra"; those that reflected on principles were seen as "losers who couldn't do the work."[67] Yet, following a study of Paul Feyerabend and his own predilections, Smolin came to the conviction that dissent and debate on principles is now necessary in quantum gravity, where a "revolution" is required; yet string theorists, he believes, are technical craftsmen that are suited for conformism and "normal science", but not seers who naturally and openly debate foundational assumptions.[68]

The prominent arenas of string theory indeed do not exhibit a culture of open debate on issues of principle. At the *Strings 2018* meeting in Okinawa, for instance, the organizers invited questions from the participants to be discussed by a panel in a plenary session. "The large majority, or rather, the plurality of questions", chair David Gross observed, "have to do with connections between string theory and the real world, experiment." Such questions typically asked "how long can string theory survive without experimental verification?" Revealingly, most contributors decided to submit their questions anonymously, which was hardly a surprise to any of those present. The opinion leaders on stage, one after another, then unanimously denied having any concern about the perceived lack of empirical connections, to which there were no follow-up questions from the audience;[69] such goings on do not exactly suggest a safe space for open discussion.

Yet, it would be wrong to follow Smolin in his characterization of string theorists as not interested in the foundations or principles of physics: subjects as holography, the information paradox or indeed the multiverse do obviously address issues of 'principle'. String theorists may prefer calculation over foundational reflection, though—while Smolin's own set is quite at home in the latter.[70]

With this difference in mind, it cannot be surprising that different versions of what kind of quantum gravity researcher is needed are articulated, and that different aspects of Einstein's biography are highlighted. The latter are both more or less grounded in the historical Einstein—but their difference illustrates the larger point: they signal, as articulations and demarcations of scholarly identities and exemplars, a Kuhnian episode of crisis and possibly paradigmatic innovation. Paying attention to these articulations will illustrate how differences of practice are translated into differences of epistemology. Insight into the debate regarding the role of the empirical—and its stalemate—may then come from historiographical appreciation of the various cultures of theory.

---

[66] Schweber (1986); Kaiser (2002; 2005; 2011); Camilleri (2009).
[67] Smolin (2006a), p. 312.
[68] See Smolin (2006a), chapters 16-20 (pp. 261-351).
[69] Strings 2018, 29 June 2018, Okinawa, https://www.youtube.com/watch?v=PlbGhv0nmJ8 (session title "50th anniversary"; consulted on 24 September 2020).
[70] See e.g. Smolin (2006b) for a philosophically informed discussion on 'background independence'.

**And yet, philosophy**

The example of Einstein is a reminder that there is an ontological dimension to the debate on what is the proper form of rationality. Einstein's criticism of the quantum theory by way of the EPR argument was linked to what he believed to be real and fundamental; this, in turn, informed what he considered to be a proper explanation and 'result'; and all these positions were reflected in his practice of unified field theory.[71] The point is: there is a link between belief about what is in the world, and what counts as explanation. Furthermore, and more directly on topic: what one considers to be in the world also informs what one considers to be 'empirical'.

And what may be considered 'empirical'? As we saw earlier in the case of testability, we again observe that physicists may have very different appreciations of what may be considered 'empirical'. For example, noted string theorist Eva Silverstein (2019, p. 373) emphasizes that string theory "participates in empirical science in several ways." Namely, it suggests constraints on early universe cosmological model building, has inspired certain scenarios for dark energy and inflation, uses "thought experiments", and produces spin-offs such as supersymmetry, which may in principle yet be testable. To many, however, these will be quite heterodox examples of what may constitute "empirical science".

Yet, even while arguing that string theory does contribute to empirical science, Silverstein also suggests that the epistemic merit of the empirical is over-rated:

> Even in empirically established theories, we empirically test only a set of measure zero of their predictions. A plethora of such tests can provide compelling evidence for a theory, which makes further predictions beyond those explicitly tested. It is not ever the case that *all* of a theory's predictions are empirically verified.[72]

The point will look familiar to philosophers: the above is a version of the problem of induction (on steroids). Silverstein does not question the merit of inductive knowledge, however, and depends on it in her own arguments, for instance when claiming that "strong evidence for string theory locally [i.e. in our universe] would support its global predictions for a landscape."[73] According to her, in quantum cosmology "valuable science" is also done the other way around, from the general to the particular: by "excluding potential alternative theories based on theoretical consistency criteria" in the "vast theory space" of the string landscape.[74] The point, to this interpreter, has a distinctly Kantian echo: in order for us to be able to know the world (i.e. for it to be "internally consistent"[75]), it must have particular features ("constraints on parameter space"[76]), and these we can establish a priori. Indeed,

---

[71] Van Dongen (2010), chapter 7.
[72] Silverstein (2019), p. 373.
[73] Silverstein (2019), p. 372; she adds in a footnote an even stronger version of the same point, attributed to Andrei Linde, who would have argued that "the universe hypothesis is no more conservative than the multiverse hypothesis in the sense that both refer to physics outside of our empirical view."
[74] Silverstein (2019), pp. 367-68.
[75] Silverstein (2019), p. 367.
[76] Silverstein (2019), p. 367.

Silverstein appears to by-pass the problem of induction and to secure progress by theory by reinstating the synthetic a priori. She uses the language of information theory to do so:

> Information is maximized when the probabilities are equal for a set of outcomes, since one learns the most from a measurement in that case. The existence of multiple consistent theoretical possibilities implies greater information content in the measurements. Therefore, theoretical research establishing this (or constraining the possibilities) is directly relevant to the question of what and how much are learned from the data.[77]

In other words, *a priori* adding (i.e. positing "multiple consistent theoretical possibilities") or excluding theories and universes ("constraining the possibilities") *expands our knowledge of the world* (that is: amounts to a synthetic judgment) because they alter the probability distribution of (in principle) possible "outcomes". This can be done rationally by insisting on virtues like, particularly, "consistency". This, then, leads Silverstein to conclude that, e.g., "the landscape has led to new empirical [or, rather: synthetic] information about the early universe and provided for a consistent interpretation of the dark energy."[78]

Even though in his early career Einstein disliked Kant, he was aware of and explicit about the Kantian nature of his later unification project. String theorists, however, not so much. Some may even be uncomfortable to find themselves identified with a Kantian agenda: Kant, after all, is quite a 'philosopher', and no one appears happy to be seen to do 'philosophy'. Yet, that sentiment only highlights that our disciplinary demarcations are primarily a consequence of institutional labeling: they do not reflect any intrinsic epistemic necessity. Indeed, it seems at least as valid to claim that Kantian philosophy was an essential element to the articulation of the law of energy conservation[79] as to argue, with Silverstein, that studies of the landscape hypothesis contribute to empirical cosmology. In light of that comparison, the Kantian nature of string theory may actually be construed as an argument in favor of its practice. Still, Silverstein's playing up of the empirical credentials of string theory— while deemphasizing the necessity of the role of the empirical—serves primarily the purpose of arguing for string theory *as a subject of physics*; keeping a distance from philosophy is part of that effort.

Indeed, Joseph Polchinksi described himself as a "practical physicist" who does not "use expressions like 'post-empirical'"; not exactly partial to philosophical reflection, he added that if only the Planck scale were attainable at LHC levels, "we would not be sitting around here whining about falsifiability." He added that restraining physics to scales accessible to "direct observation" would render science "too weak"; it would be "to decree that some aspects of the natural world are outside of its domain."[80] Polchinski, too, believed that string theory did relate to the empirical, in particular since "only the multiverse made […] the prediction most worth making" in cosmology—that the universe's vacuum energy was finite and non-zero, as (arguably) observed in 1997.[81]

---

[77] Silverstein (2019), p. 368.
[78] Silverstein (2019), p. 371.
[79] On Kant's influence in the case of energy conservation, see e.g. Heimann (1974).
[80] Polchinksi (2019a), p. 340.
[81] Polchinski (2019a), p. 348.

Polchinski referred for this prediction particularly the review article on "The cosmological constant problem" by Stephen Weinberg from 1989, which briefly discussed inflationary multiverse ideas among a number of "Anthropological Considerations" that support a cosmological constant reasonably close to the currently accepted number.[82] Yet, that discussion predates modern string theory versions of the multiverse: so to frame Weinberg's account as a prediction articulated by the latter appears at face value as an example of rather Whiggish historiography. More fundamental, however, is the question whether any multiverse account can be seen as "predictive" in a sense that is familiar in physics: in this account of cosmology, the laws of nature of our universe have become contingent and local, ultimately only picked out among an infinitude of universes, laws and parameters because these laws and parameters allow for life as we know it: the coincidence of the environment that we live in then determines the laws, rather than that the laws determine the environment.[83] When following through on that logic, simply observing that we find ourselves in our life supporting version of the universe can then be crafted as an empirically meaningful statement.[84] One might object, however, that the parallel universes invoked to explain the existence of ours are beyond our ability to observe: they are inaccessible in principle (Ellis 2017), which can hardly reinforce the empiricist credentials of the multiverse scenario.

Yet, no single string theorist will deny the ultimate authority of observation in empirical science.[85] In fact, string theory originated in the S-matrix program for nuclear interactions: this program was principle-driven but still depended strongly on empirical input,[86] just like the broader particle physics tradition in which it has always been embedded. This dependence changed when the theory began to be considered as a theory of quantum gravity. Yet we currently do not have a full historical-philosophical grasp of how that affected the implicit epistemological positions of the subject, and how these developed over time.[87] Yet, such historical understanding will be essential to see how its positions today regarding the role of the empirical came about, and thus in understanding the current differences of opinion: more historiography of string theory is needed.

To develop historical understanding of what is considered empirical by string theorists, one also needs to develop a sense of the subject's ontology, and the history thereof. Today, for example, linking semiclassical gravity in the AdS space of the AdS/CFT correspondence to "physics that is measurable on our planet" may be achieved in condensed matter systems, in the words of a group in Leiden that works on this subject.[88] Yet, that raises immediate philosophical questions: since the two descriptions, valid in different regimes, are related by a *duality*, what does a description in terms of one theory actually tell us about the life of objects in some 'real' world caught by the other description? For example, it has been suggested that in the AdS/CFT duality, one side of the duality can be seen as being emergent from the other side—how would that impact the sense of description across the duality?[89] And how would that change in case of an identity relation across the duality? The important point is: answers to such ontological questions impact on *whether* dualities explain, *what* they are in fact explaining, and *how* those explanations *relate to the empirical world*. Furthermore, these answers may differ between communities.

---

[82] Weinberg (1989), pp. 6-9; on p. 7, Weinberg mentioned that "[i]n a model of Linde [...], fluctuations in scalar fields produce exponentially expanding regions of the universe, within which further fluctuations produce further subuniverses, and so on. Since these subuniverses arise from fluctuations in the fields, they have differing values of various 'constants' of nature." This brief discussion of multiverse ideas was included as one of a number of scenarios that Weinberg discussed as anthropic accounts.

[83] Kragh (2009), in particular p. 540.

[84] String theorist Robbert Dijkgraaf made this point in a recent public debate on the empirical nature of string theory ("Bèta break", University of Amsterdam, 28 November 2017).

[85] Yet, like Einstein, they may not insist on its necessity in creatively formulating theories: see his "On the method" (1933).

[86] Cappelli et al. (2012); Rickles (2014). It should be noted, however, that in 1970 Fermilab terminated all but one of its novel theory group members after only less than two years when the group gravitated from phenomenology to dual models and thus away from experiment; see Hoddeson et al. (2008), pp. 139-141.

[87] See however Castellani (2019) for a brief account of early string theory that emphasizes the role of theoretical virtues.

Opinions about what is 'real' will affect which argument or calculation convinces. When asking ontological questions, it will not be enough to simply state, as Polchinksi did, that string theory "exists" and that e.g. the black hole microstates "can be understood" in string theory, via, for example, the Strominger-Vafa analysis.[90] Even though this example is often cited as a success of string theory, its appreciation depends on how one assesses the underlying duality-type relationship: are the two systems—the microscopic D-brane system and the thermodynamic black hole—truly *the same*? Or is the numerical identity between entropies simply an accidental consequence of the mathematical peculiarities of the theories involved? After all, the entropy counting in the D-brane system is performed with the gravitational force turned off. Can one then legitimately speak of a *black hole*?[91] Perhaps unsurprisingly, string critic Lee Smolin is willing to entertain a "pessimistic point of view" on the latter point.[92] So, to assess what is rationally considered evidence in string theory, one needs to address—and at times assist in articulating—the ontological positions of its practitioners (and, in turn, those of its critics). Fortunately, much work has been done in recent years by philosophers on particularly the interpretation of duality, which can inform the string theory debate on exactly these issues.[93] This will aid historiographical reconstruction of string theory's implicit epistemology.

Ontological conviction about what is most fundamental in our description of the world is tied to practice: in the black hole information paradox, for example, it can hardly be a surprise that traditionally many relativists have been much more inclined to go along with Hawking's original semi-classical analysis of information loss, grounded in the causal structure of the black hole spacetime, than the field theorists' insistence on quantum unitarity.[94] In this sense, identity and ontological conviction matter: they condition positions on explanation and understanding. Thus, Joseph Polchinksi expressed surprise that Ellis and Silk do not agree with the multiverse scenario, and are apparently content to leave the value of the cosmological constant up to measurement,[95] without any further need for 'explanation'. An important element for Polchinski's' own argument for the multiverse is in fact autobiographical—his personal struggle with Weinberg's anthropic reasoning and eventually his own realization of its possibility in string theory. In the end, as we saw, many of Polchinski's arguments for string theory *as physics*, are arguments that articulate what it means to be a good physicist; that express what kind of practice is epistemically virtuous.

---

[88] See: https://www.lorentz.leidenuniv.nl/zaanen/wordpress/research/anti-de-sittercondensed-matter/, cited on 17 September 2020; for an overview of the subject see their book: Zaanen et al. (2015); earlier contributions to the subject include Hartnoll et al. (2007, 2008); strong criticism of the general applicability of the AdS/CFT correspondence to condensed matter systems has also been expressed by one of the latter's leading scholars; see Anderson (2013).
[89] Dieks et al. (2015).
[90] Polchinksi (2019a), pp. 349, 346 respectively; Strominger and Vafa (1996).
[91] For a discussion of particularly this point, see Van Dongen et al. (2020), pp. 115-117; see also De Haro et al. (2020) for more on the Strominger-Vafa analysis from an HPS perspective.
[92] Smolin (2006a), pp. 139-141.
[93] See e.g. Rickles (2011, 2013); Matsubara (2013); Dieks et al. (2015); Castellani (2017); Castellani and De Haro (2020); De Haro and Butterfield (2019); De Haro et al. (2017); De Haro (2019, 2020), Read (2016).

**Conclusion**

Polchinski and Silverstein are opinion-leaders in string theory, and have been among the first to attempt to articulate the epistemic practice of their subject; this is important and immensely useful for philosophers. Their standards of explanation and what exactly constitutes the 'empirical' have been shown above to differ considerably from those of string theory's critics. Yet, heterodox opinions of what may be considered 'empirical' and 'predictive' should not be taken to signal that these labels are unjustly applied, but simply that a substantial group of scholars, who emphatically do consider themselves physicists, consider certain types of argument as epistemically relevant to, and valid expressions of empirical science, while others will disagree. These judgments are expressions of different cultures of rationality, rooted in different practices of theory: namely, in the use of string theory, or its absence. This does suggest that discussions about normative methodological prescriptions—and this has been the main thrust of this paper—will probably not be resolved and may not even substantially inform the stalemate in the current controversy. Yet, as I have tried to argue, historiography, when sufficiently sensitive to string theory's implicit epistemology, might.

Lee Smolin (2006a, pp. 301-303) has stated that democratic dissent on the basis of shared and publicly available evidence is a necessary element of science. What the preceding section has shown, however, is that what is construed as evidence is inherently conditioned by disciplinary culture; and that it is that conditioning that drives dissent in the first place. Indeed, all participants in the debate on string theory are convinced that they uphold proper standards of rationality, yet disagree on what those are. The contentious and even emotional nature of the debate is ultimately due to its personal nature, exactly because at stake is what it means to be a physicist: different positions in the crisis in fundamental physics are linked to different senses of virtue and identity—and consequently even different versions of Einstein.

---

[94] Van Dongen and De Haro, forthcoming; talks at EPSA 2019, Geneva; 't Hooft meeting, Utrecht, 2019; Information paradox philosophy workshop, Cambridge, 2019; GGI workshop non-empirical physics, Florence, 2019; see further e.g. Van Dongen and De Haro (2004); Susskind (2008).

[95] At least, that is how Polchinksi (2019b, p. 362) interprets their mention of unimodular gravity as alternative to a multiverse account.

The sentiment expressed by string theorists, whatever their reasoning, is that there can be explanatory power in exclusively theoretical argumentation. That point has been developed by Dawid (2013, 2017), who argues that theories that provide 'unexpected explanations of additional phenomena', or theories for which 'no alternative' can be conceived, particularly when those theories are built on methods that have proven successful in the past, do deserve additional epistemic confidence. These epistemic warrants apply to string theory, Dawid claims, and should be given proper authority as they function as such in the field: in his opinion, the sustained prominence of string theory entails that philosophers of science need to revise their traditional accounts of confirmation and accept such novel non-empirical norms.

Yet, citing Kuhn, I have made the case that such norms need not be considered so very novel; and more importantly, that scientific rationality is not captured by reductive accounts of confirmation, but is rather more akin to something like reasoned judgement on the basis of incomplete information; that principles and maxims are weighed, like values are weighed in moral judgment. Whatever value then prevails is strongly related to what ideal circulates of how a scholar should be. Such ideals and the values that they represent are furthermore shared socially: they are inherently social. Yet, some scholars express frustration at "social […] biases" they observe in the assessment of theory today (Oriti 2019, p. 149; Hossenfelder 2017)—yet, most recent historiography will insist that criteria of rationality are inherently social: values only have force *if they are shared*. Indeed, shared norms of assessment and principle are exactly what build a discipline.

Demarcation and its criteria have a social function: they function to exclude certain practices and people from the discipline proper—qualifying a research tradition as non-scientific, or to be "rather regarded as philosophy" (Ellis 2017, p. 33), does exactly that. Yet the vast majority working in the tradition of string theory are in fact employed by physics departments.

The crisis in physics is both inextricably social and epistemic: it is a crisis of identity. So here is a resolution. Let us accept that this is an example of incommensurability, and embrace the need to fashion a novel identity in what is still "a no-man's land between mathematics, physics and philosophy" (Ellis and Silk 2014, p. 321). That is what Einstein tried to do for his unified field theory work, and Galileo achieved when he secured his job as court *philosopher* in Florence. In this sense, I do agree with Richard Dawid that philosophers—and physicists—should accept that there has been a "shift of the scientific paradigm" (Dawid 2006, p. 301). I further would like to propose a name for the new non-empirical paradigm, honoring its Kantian credentials: let's call it '*meta*'-physics, and keep funding it as generously as before.

**Acknowledgments** I am most grateful to Alex Blum and other organizers of the workshop 'Non-Empirical Physics from a Historical Perspective' in March 2019 at the Max Planck Institute for the History of Science in Berlin for putting together an exciting event; and equally grateful for the editors of this special issue for their solid work and patience. I thank Jeremy Butterfield, Elena Castellani, Alejandra Castro, Richard Dawid, Sjang ten Hagen, Sebastian de Haro, David Kaiser, Robert van Leeuwen, Friedrich Stadler, Manus Visser and the journal's referees for helpful feedback; just as I am indebted to audiences at the Austrian Academy of Science & the Vienna Circle Institute, the GGI in Florence, and the Dutch National Institute for Subatomic Physics in Amsterdam for insightful debate.

**References**

Abi, B., et al. (Muon g-2 collaboration), (2021). "Measurement of the positive muon anomalous magnetic moment to 0.46 ppm", *Physical Review Letters*, 126, 141801.

Anderson, P.W. (2013). "Strange connections to strange metals", *Physics Today*, 66, 9.

Arabatzis, T. (2006). "On the inextricability of the context of discovery and the context of justification", pp. 215-230 in J. Schickore and F. Steinle (eds.), *Revisiting Discovery and Justification*, Dordrecht: Springer.

Avin, S. (2019). "Centralised funding and epistemic exploration", *The British Journal for the Philosophy of Science*, 70, 629-656.

Bellon, R. (2015). *A Sincere and Teachable Heart: Self-Denying Virtue in British Intellectual Life, 1736-1859*. Leiden: Brill.

Bertone, G. and T. Tait (2018). "A new era in the search for dark matter", *Nature*, 562, 51-56.

Beyerchen, A.D. (1977). *Scientists under Hitler. Politics and the physics community in the Third Reich*. New Haven: Yale University Press.

Bezuidenhout, L., E. Ratti, N. Warne and D. Beeler (2019). "Docility as a primary virtue in scientific research", *Minerva*, 57, pp. 67-84.

Biagioli, M. (1993). *Galileo, courtier. The practice of science in the culture of absolutism*. Chicago: The University of Chicago Press.

Bird, K. and M.J. Sherwin (2006). *American Prometheus: The triumph and tragedy of J. Robert Oppenheimer*, New York: Vintage.

Blom, Th. and M. Wessel (2017). "Deze wetenschappers strijden tegen alternatieve feiten", *De Volkskrant*, 29 April 2017.

Butterfield, J. (2019). "Lost in math? Essay review of Sabine Hossenfelder, 'Lost in math: How beauty leads physics astray'", *Physics in Perspective*, 21, 63-86. arXiv:1902.03480

Camilleri, K. (2009). "A history of entanglement: Decoherence and the interpretation problem", *Studies in History and Philosophy of Science*, Part B: *Studies in History and Philosophy of Modern Physics*, 40, 290-302.

Cappelli, A., E. Castellani, F. Colomo, and P. di Vecchia (eds.) (2012). *The birth of string theory*, Cambridge: Cambridge University Press.

Castellani, E. (2017). "Duality and 'particle' democracy", *Studies in History and Philosophy of Science*, Part B: *Studies in History and Philosophy of Modern Physics*, 59, 100-108.

Castellani, E. (2019). "Scientific methodology: A view from early string theory", pp. 173-183 in R. Dardashti, R. Dawid and K. Thébault (eds.), *Why trust a theory? Epistemology of fundamental physics*, Cambridge: Cambridge University Press.

Castellani, E. and S. De Haro (2020). "Duality, fundamentality and emergence", pp. 195-216 in: D. Glick, G. Darby and A. Marmodoro (eds.), *The foundation of reality: fundamentality, space and time*, Oxford: Oxford University Press. arXiv:1803.09443

Dardashti, R. and S. Hartmann (2019). "Assessing scientific theories. The Bayesian approach", pp. 67-83 in: R. Dardashti, R. Dawid and K. Thébault (eds.), *Why trust a theory? Epistemology of fundamental physics*, Cambridge: Cambridge University Press.

Daston, L. (1995). "The moral economy of science", *Osiris*, pp. 2-24.

Daston, L. and P. Galison (2007). *Objectivity*. New York: Zone Books.

Daston, L. and H.O. Sibum. (2003). "Introduction: Scientific personae and their histories", *Science in Context*, 16, 1-8.

Dawid, R. (2006). "Underdetermination and theory succession from the perspective of string theory", *Philosophy of Science*, 73, 298-322.

Dawid, R. (2013). *String theory and the scientific method*. Cambridge: Cambridge University Press.

Dawid, R. (2019). "The significance of non-empirical confirmation in fundamental physics", pp. 99-119 in R. Dardashti, R. Dawid and K. Thébault (eds.), *Why trust a theory? Epistemology of fundamental physics*, Cambridge: Cambridge University Press.

Dawid, R. (2020). "Meta-empirical confirmation: Addressing three points of criticism", http://philsci-archive.pitt.edu/16908/.

Dawid, R., S. Hartmann, J. Sprenger (2015). "The no alternatives argument", *The British Journal for the Philosophy of Science*, 66, 213-234.

De Haro, S. (2019). "The heuristic function of duality", *Synthese*, 196, 5169-5203. arXiv:1801.09095.

De Haro, S. (2020). "On empirical equivalence and duality", pp. 91-106 in S. de Bianchi and C. Kiefer (eds.), *One hundred years of gauge theory*, Berlin: Springer, arXiv:2004.06045.

De Haro, S., and J. Butterfield (2019). "On symmetry and duality", *Synthese*, http://philsci-archive.pitt.edu/16093/

De Haro, S., N. Teh, and J. Butterfield (2017). "Comparing dualities and gauge symmetries", *Studies in History and Philosophy of Science*, Part B: *Studies in History and Philosophy of Modern Physics*, 59, 68–80. http://philsci-archive.pitt.edu/12009.

De Haro, S., J. van Dongen, M. Visser, and J. Butterfield (2020). "Conceptual analysis of black hole entropy in string theory", *Studies in History and Philosophy of Science*, Part B: *Studies in History and Philosophy of Physics*, 69, 82-111. http://philsci-archive.pitt.edu/15883/.

De Swart, J., G. Bertone, and J. van Dongen (2017). "How dark matter came to matter", *Nature Astronomy*, 1 0059. arXiv:1703.00013.

De Vries, J. (2021). "Voorzichtige opwinding op CERN over afwijking standaardmodel", *Nederlands Tijdschrift voor Natuurkunde*, 87(5), 8-9.

De Waal, E. and S.L. ten Hagen (2020). "The concept of fact in German physics around 1900: A comparison between Mach and Einstein", *Physics in Perspective*, 22, 55-80.

Dieks, D., J. van Dongen, and S. De Haro (2015). "Emergence in holographic scenarios for gravity", *Studies in History and Philosophy of Science* Part B: *Studies in History and Philosophy of Modern Physics*, 52, 203-26. arXiv:1501.04278.

Einstein, A. ([1918] 1994). "Principles of research", pp. 244-248 in *Ideas and Opinions*, New York: The Modern Library.

Einstein, A. ([1930] 1994). "Religion and science", pp. 39-43 in *Ideas and Opinions*, 39-43, New York: The Modern Library.

Einstein, A. ([1933] 1994). "On the method of theoretical physics", pp. 296-303 in *Ideas and Opinions*, New York: The Modern Library.

Einstein, A. ([1934] 1994). "Good and evil", p. 13 in *Ideas and Opinions*, New York: The Modern Library.

Einstein, A. ([1948] 1994). "Religion and science: Irreconcilable?", pp. 53-57 in *Ideas and Opinions*, New York: The Modern Library.

Einstein, A. ([1949] 1997). "Autobiographisches [and translation: Autobiographical Notes]", pp. 1-94 in P.A. Schilpp (ed.), *Albert Einstein: Philosopher-Scientist*, La Salle: Open Court.

Einstein, A. (1950). "On the generalized theory of gravitation", *Scientific American* 182, 13-17.

Ellis, G. (2017). "The domain of cosmology and the testing of scientific theories", pp. 3-39 in K. Chamcham, J. Silk, J.D. Barrow and S. Saunders (eds.), *The Philosophy of Cosmology*, Cambridge: Cambridge University Press.

Ellis, G. and J. Silk (2014). "Scientific method: Defend the integrity of physics", *Nature*, 516, 321-323.

Forman, P. (1984). "Il Naturforscherversammlung a Nauheim del settembre 1920: una introduzione alla vita scientifica nella repubblica di Weimar", pp. 59–78 in G. Battimelli, M. de Maria, and A. Rossi (eds.), *La ristrutturazione delle scienze tra le due guerre mondiali*, Rome: La goliardica.

Galison, P. (1995). "Theory bound and unbound: Superstrings and experiment", pp. 369-408 in F. Weinert, *Laws of nature. Essays on the philosophical, scientific and historical dimensions*. Berlin: de Gruyter.

Galison, P. (2004). "Mirror symmetry: persons, values and objects", pp. 23-63 in M. Norton Wise (ed.), *Growing explanations: Historical perspectives on recent science*. Durham NC: Duke University Press.

Garisto, D. (2021). "Long awaited muon measurement boosts evidence for new physics", *Scientific American*, 7 April 2021.

Gefter, A. (2005). Is string theory in trouble? *New Scientist*, December issue.

Gilbert, M.K., and A. Loveridge (2020). "Subjectifying objectivity: Delineating tastes in theoretical quantum gravity research", *Social Studies of Science*, https://doi.org/10.1177/0306312720949691.

Goenner, H.F.M. (2004). "On the history of unified field theories", *Living Reviews in Relativity*, 7(2).

Goenner, H.F.M. (2014). "On the history of unified field theories, Part II. (ca. 1930—ca.1965)", *Living Reviews in Relativity*, 17(5).

Greene, B. (1999). *The elegant universe. Superstrings, hidden dimensions, and the quest for the ultimate theory*. New York: W.W. Norton.

Guldi, J. and D. Armitige (2014). *The history manifesto*. Cambridge: Cambridge University Press.

Guth, A., D.I. Kaiser, A.D. Linde, Y. Nomura, G.R. Bond, F. Bouchet, S. Carroll, G. Esthathiou, S. Hawking, R. Kallosh, E. Komatsu, L. Krauss, D.H. Lyth, J. Maldacena, J.C. Mather, H. Peiris, M. Perry, L. Randall, M. Rees, M. Sasaki, L. Senatore, E. Silverstein, G.F. Smoot, A. Starobinsky, L. Susskind, M.S. Turner, A. Vilenkin, S. Weinberg, R. Weiss, F. Wilczek, E. Witten, and M. Zaldariagga (2017), "A cosmic controversy", *Scientific American* 316, https://blogs.scientificamerican.com/observations/ a-cosmic-controversy/ (downloaded 1 July 2018).

Hamilton, J.-C. (2014). "What have we learned from observational cosmology?" *Studies in History and Philosophy of Science,* Part B: *Studies in History and Philosophy of Modern Physics* 46 (2014), 70-85.

Hartnoll, S.A., P.K. Kovtun, M. Müller, and S. Sachdev (2007). "Theory of the Nernst effect near quantum phase transitions in condensed matter and dyonic black holes", *Physical Review B*, 76, 144502.

Hartnoll, S.A., C. Herzog, and G. Horowitz (2008). "Building a holographic superconductor", *Physical Review Letters*, 101, 31601.

Heilbron, J. (1992). "Creativity and big science", *Physics Today*, 45 (11), pp. 42-47.

Heilbron, J. (2010). *Galileo*. Oxford: Oxford University Press.

Heimann, P.M. (1974). "Helmholtz and Kant: The metaphysical foundations of *Über die Erhaltung der Kraft*", *Studies in History and Philosophy of Science*, 5, pp. 205-238.

Heisenberg, W. (1989). *Encounters with Einstein and other essays on people, places and particles*. Princeton: Princeton University Press.

Hicks, D.J. and T.A. Stapleford (2016). "The virtues of scientific practice: MacIntyre, virtue ethics, and the historiography of science", *Isis*, 107, pp. 449-472.

Higgitt, R. (2007). *Recreating Newton: Newtonian biography and the making of nineteenth century history of science*, London: Pickering and Chatto.

Hoddeson, L., and A.W. Kolb, and C. Westfall (2008). *Fermilab: Physics, the frontier and megascience*. Chicago: The University of Chicago Press.

Holton, G. (1988). *Thematic origins of scientific thought. Kepler to Einstein*. Cambridge MA: Harvard University Press.

Hossenfelder, S. (2017). "Science needs reason to be trusted", *Nature Physics*, 13, 316-317.

Hossenfelder, S. (2018). *Lost in math. How beauty leads physics astray*. New York: Basic Books.

Howard, D. (1994). "Einstein, Kant, and the origins of logical empiricism", pp. 45-105 in W. Salmon and G. Wolters (eds.), *Language, Logic and the Structure of Scientific Theories*, Pittsburgh: University of Pittsburgh Press.

Ijjas, A., P. Steinhardt, and A. Loeb (2017). "Pop goes the universe", *Scientific American* 316(2), 32-39.

Illy, J. (2019). "Einstein's gyros", *Physics in Perspective*, 21, pp. 274-295.

Janssen, M., J. Norton, J. Renn, T. Sauer and J. Stachel, (2007). *The genesis of general relativity*, vols. 1 & 2. *Einstein's Zurich notebook*. Dordrecht: Springer.

Janssen, M. and J. Renn. (2007). "Untying the knot: How Einstein found his way back to the field equations discarded in the Zurich notebook", pp. 839-925 in: Janssen et al. (2007).

Jones, M.L. (2006). *The Good Life in the Scientific Revolution: Descartes, Pascal, Leibniz, and the Cultivation of Virtue*. Chicago: The University of Chicago Press.

Jungnickel, J. and R. McCormmach (1986). *Intellectual mastery of nature. Theoretical Physics from Ohm to Einstein*. 2 vols. Chicago: The University of Chicago Press.

Kaiser, D.I. (2002). "Cold war requisitions, scientific manpower and the production of American physicists after World War II", *Historical Studies in the Physical and Biological Sciences* 33, 131-159.

Kaiser, D.I. (2005). *Drawing theories apart. The dispersion of Feynman diagrams in postwar physics*. Chicago: The University of Chicago Press.

Kaiser, D.I. (2011). *How the hippies saved physics. Science, counterculture and the quantum revival*. New York: W.W. Norton.

Kidd, I.J. (2014). "Was Sir William Crookes epistemically virtuous?", *Studies in History and Philosophy of Science* Part C: *Studies in History and Philosophy of Biology and Biomedical Sciences*, 48A, pp. 67-74.

Kox, A.J. (1988). "Hendrik Antoon Lorentz, the ether, and the general theory of relativity", *Archive for History of Exact Sciences* 38, pp. 67–78.

Kox, A.J. (2008). *The Scientific Correspondence of H.A. Lorentz, Volume I*. New York: Springer.

Kox, A.J. (2019). *Hendrik Antoon Lorentz, natuurkundige 1853-1928. 'Een levend kunstwerk'*. Amsterdam: Balans.

Kragh, H. (1999). *Quantum generations. A history of physics in the twentieth century*. Princeton: Princeton University Press.

Kragh, H. (2009). "Contemporary history of cosmology and the controversy over the multiverse", *Annals of Science*, 66, 529-552.

Kragh, H. (2014). "Testability and epistemic shifts in modern cosmology", *Studies in History and Philosophy of Science*, Part B: *Studies in History and Philosophy of Modern Physics*, 46, pp. 48-56.

Kragh, H. (2019). "Fundamental theories and epistemic shifts: Can history of science serve as a guide?", pp. 13-28 in R. Dardashti, R. Dawid and K. Thébault (eds.), *Why trust a theory? Epistemology of fundamental physics*, Cambridge: Cambridge University Press. arXiv:1702.05648 [physics.hist-ph].

Krige, J. (2006). *American hegemony and the postwar reconstruction of science in Europe.* Cambridge MA: MIT Press.

Krige, J. (2016). *Sharing knowledge, shaping Europe. US technological collaboration and nonproliferation*. Cambridge MA: MIT Press.

Kuhn, T. (1962). *The structure of scientific revolutions*, Chicago: The University of Chicago Press.

Kuhn, T. (1977a). "Logic of discovery or psychology of research?", pp. 266-292 in *The essential tension*, Chicago: The University of Chicago Press.

Kuhn, T. (1977b). "Objectivity, value judgment, and theory choice", pp. 320-339 in *The essential tension*, Chicago: The University of Chicago Press.

Linde, A. (2015). "A brief history of the multiverse", arXiv:1512.01203 [hep-th].

Maldacena, J. (1998). "The large N limit of superconformal field theories and supergravity", *Advances in Theoretical and Mathematical Physics*, 2, 231–252.

Matsubara, K. (2013). "Realism, underdetermination and string theory dualities", *Synthese*, 190, 471-489.

Merritt, D. (2017). "Cosmology and convention", *Studies in History and Philosophy of Science*, Part B: *Studies in History and Philosophy of Modern Physics*, 57, 41-52.

Norton, J. (2000). "'Nature is the realization of the simplest conceivable mathematical ideas': Einstein and the canon of mathematical simplicity", *Studies in History and Philosophy of Science*, Part B: *Studies in History and Philosophy of Modern Physics*, 31, pp. 135-170.

Oriti, D. (2019). "No alternative to proliferation", pp. 125-153 in R. Dardashti, R. Dawid and K. Thébault (eds.), *Why trust a theory? Epistemology of fundamental physics*, Cambridge: Cambridge University Press.

Overbye, D. (2017), "Yearning for new physics at CERN, in a post-Higgs way", *New York Times*, 26 June 2017.

Overbye, D. (2021), "A tiny particle's wobble could upend the known laws of physics", *New York Times*, 7 April 2021.

Pais, A. (1982). *'Subtle is the Lord…' The science and life of Albert Einstein*. Oxford: Oxford University Press.

Paul, H. (2014). "What is a scholarly persona? Ten theses on virtues, skills, and desires", *History and Theory* 53, 348–371.

Pennock, R.T. (2019). *An Instinct for Truth: Curiosity and the Moral Character of Science*. Cambridge, MA: MIT Press.

Polchinksi, J. (2019a). "String theory to the rescue", pp. 339-353 in R. Dardashti, R. Dawid and K. Thébault (eds.), *Why trust a theory? Epistemology of fundamental physics*, Cambridge: Cambridge University Press; arXiv:1512:02477 [hep-th].

Polchinksi, J. (2019b). "Why trust a theory? Some further remarks", pp. 354-364 in R. Dardashti, R. Dawid and K. Thébault (eds.), *Why trust a theory? Epistemology of fundamental physics*, Cambridge: Cambridge University Press; arXiv:1601:06145 [hep-th].

Read, J. (2016), "The interpretation of string-theoretic dualities", *Foundations of Physics*, 46, 209-235.

Renn, J. and Sauer, T. (2007). "Pathways out of classical physics. Einstein's double strategy in is search for the gravitational field equations", pp. 113-312 in Janssen et al. (2007).

Rickles, D. (2011). "A philosopher looks at string dualities", *Studies in History and Philosophy of Science*, Part B: *Studies in History and Philosophy of Modern Physics*, 42, 54-67.

Rickles, D. (2012). "AdS/CFT duality and the emergence of spacetime", *Studies in History and Philosophy of Science*, Part B: *Studies in History and Philosophy of Modern Physics*, 44, 312–320.

Rickles, D. (2014). *A brief history of string theory. From dual models to M-theory*. Heidelberg: Springer.

Ritson, S. and K. Camilleri (2015). "Contested boundaries: The string theory debates and ideologies of science", *Perspectives on Science*, 23, 192-227.

Rovelli, C. (2019). "The dangers of non-empirical confirmation", pp. 120-124 in R. Dardashti, R. Dawid and K. Thébault (eds.), *Why trust a theory? Epistemology of fundamental physics*, Cambridge: Cambridge University Press.

Rowe, D. (2012). "Einstein and relativity: What price fame?" *Science in Context*, 25, 197-246.

Saarloos, L. (2017). "Virtues of courage and virtues of restraint: Tyndall, Tait and the use of imagination in late Victorian science", pp. 109-128 in: J. van Dongen and H. Paul (eds.) *Epistemic Virtues in the Sciences and the Humanities*, Boston Studies in the Philosophy and History of Science (vol. 321), Springer: 2017.

Saarloos, L. (2021). *The scholarly self under threat: Language of vice in British scholarship (1870-1910)*. PhD dissertation, Leiden University.

Salmon, W.C. (1990). "Rationality and objectivity in science, or Tom Kuhn meets Tom Bayes", pp. 175-204 in C.W. Savage (ed.), *Scientific Theories*, Minneapolis: University of Minnesota Press.

Sample, I. (2021). "Cern experiment hints at a new force of nature", *The Guardian*, 23 March 2021.

Sauer, T. (2006). "Field equations in teleparallel spacetime: Einstein's *Fernparallelismus* approach toward unified field theory", *Historia Mathematica*, 33, pp. 399-439.

Sauer, T. (2014). "Einstein's unified field theory program", pp. 281-305 in M. Janssen and C. Lehner (eds.), *The Cambridge Companion to Einstein*, Cambridge: Cambridge University Press.

Schönbeck, C. (2000). "Albert Einstein und Philipp Lenard. Antipoden im Spannungsfeld von Physik und Zeitgeschichte", *Schriften der mathematisch-naturwissenschaftliche Klasse der Heidelberger Akademie der Wissenschaften*, 8, 1-42.

Schweber, S.S. (1986). "The empiricist temper regnant: Theoretical physics in the United States, 1920-1960", *Historical Studies in the Physical and Biological Sciences*, 35, 67-93.

Schweber, S.S. (2008). *Einstein and Oppenheimer: The meaning of genius*. Cambridge, MA: Harvard University Press.

Seth, S. (2010). *Crafting the quantum: Arnold Sommerfeld and the practice of theory, 1898-1926*. Cambridge MA: MIT Press.

Shapin, S. (1994). *A social history of truth. civility and science in seventeenth-century England*. Chicago: The University of Chicago Press.

Shapin, S. (2008). *The scientific life. A moral history of a late modern vocation.* Chicago: The University of Chicago Press.

Silverstein, E. (2017). "The dangerous irrelevance of string theory", pp. 365-376 in R. Dardashti, R. Dawid and K. Thébault (eds.), *Why trust a theory? Epistemology of fundamental physics*, Cambridge: Cambridge University Press; arXiv:170602790 [hep-th].

Smith, C. (1998). *The science of energy: A cultural history of energy physics in Victorian Britain*, London: The Athlone Press.

Smolin, L. (2006a). *The trouble with physics. The rise of string theory, the fall of a science, and what comes next*. Boston: Houghton Mifflin.

Smolin, L. (2006b). "The case for background independence", pp. 196-239 in: D. Rickles, S. French and J. Saatsi (eds.), *The Structural Foundations of Quantum Gravity*, Oxford: Oxford University Press. arXiv:hep-th/0507235

Smolin, L. (2014). "String theory and the scientific method", *American Journal of Physics* 82, 1105-1107.

Söderqvist, T. (1997). "Virtue ethics and the historiography of science", *Danish Yearbook of Philosophy*, 32, pp. 45-64.

Staley, R. (2008). *Einstein's generation. The origins of the relativity revolution*. Chicago: The University of Chicago Press.

Stanley, M. (2017). "Religious and scientific virtues: Maxwell, Eddington and persistence", pp. 49-61 in: J. van Dongen and H. Paul (eds.) *Epistemic Virtues in the Sciences and the Humanities*, Boston Studies in the Philosophy and History of Science (vol. 321), Springer: 2017.

Strominger, A. and C. Vafa. (1996). "The microscopic origin of the Bekenstein-Hawking entropy", *Physics Letters* B379, 99-104; arXiv:hep-th/9601029.

Stump, D.J. (2007). "Pierre Duhem's virtue epistemology", *Studies in History and Philosophy of Science* Part A, 38, pp. 149-159.

Susskind, L. (2008). *The black hole war. My battle with Stephen Hawking to make the world safe for quantum mechanics*. New York: Little Brown and Company.

Tai, C. (2017). "Left radicalism and the Milky way: Connecting the scientific and socialist virtues of Anton Pannekoek", *Historical Studies in the Natural Sciences*, 47, 200-254.

Talbott, W. (2008). "Bayesian epistemology", *The Stanford Encyclopedia of Philosophy*, https://plato.stanford.edu/archives/win2016/entries/epistemology-bayesian/ (downloaded 16 October 2020).

Ten Hagen, S.L. (2019). "How facts shaped modern disciplines: The fluid concept of fact and the common origins of German physics and historiography", *Historical Studies in the Natural Sciences*, 49, 300-337.

Ten Hagen, S.L. (2021). *History and physics entangled. Disciplinary intersections in the long nineteenth century*. PhD thesis, University of Amsterdam.

Turri, J., M. Alfano, and J. Greco. (2017). "Virtue epistemology", *The Stanford Encyclopedia of Philosophy*, https://plato.stanford.edu/entries/epistemology-virtue/ (downloaded 15 July 2018).

Van Dongen, J. (2004). "Einstein's methodology, semivectors and the unification of electrons and protons", *Archive for History of Exact Sciences*, 58, pp. 219-254.

Van Dongen, J. (2007). "Reactionaries and Einstein's Fame: 'German scientists for the preservation of pure science,' Relativity and the Bad Nauheim meeting", *Physics in Perspective*, 9, 212-230. arXiv:1111.2194.

Van Dongen, J. (2010). *Einstein 's unification*. Cambridge: Cambridge University Press.

Van Dongen, J. (2012). "Mistaken identity and mirror images: Albert and Carl Einstein, Leiden and Berlin, relativity and revolution", *Physics in Perspective*, 14, 126-177. arXiv:1211.3309

Van Dongen, J. (2015). *Cold war science and the transatlantic circulation of knowledge*. Leiden: Brill.

Van Dongen, J. (2017). "The epistemic virtues of the virtuous theorist: On Albert Einstein and his autobiography", pp. 60-92 in J. van Dongen and H. Paul (eds.) *Epistemic Virtues in the Sciences and the Humanities*, Boston Studies in the Philosophy and History of Science (vol. 321), Springer: 2017. arXiv:2002.01301.

Van Dongen, J., and S. De Haro (2004). "On black hole complementarity", *Studies in History and Philosophy of Science*, Part B: *Studies in History and Philosophy of Physics*, 35, 509-525.

Van Dongen, J., S. De Haro, M. Visser and J. Butterfield. (2020) "Emergence and correspondence for string theory black holes", *Studies in History and Philosophy of Science*, Part B: *Studies in History and Philosophy of Physics*, 69, 112- 127. arXiv:1904.03234.

Van Dongen, J., and H. Paul (2017) *Epistemic virtues in the sciences and the humanities*. Boston Studies in the Philosophy and History of Science, vol. 321. Dordrecht: Springer.

Vizgin, V. (1994). *Unified field theories in the first third of the twentieth century*. Basel: Birkhäuser.

Wang, J. (2012). "Physics, emotion and the scientific self: Merle Tuve's cold war", *Historical Studies in the Natural Sciences*, 42, 341-388.

Wazeck, M. (2013). *Einstein's opponents. The public controversy about the theory of relativity in the 1920s*. Cambridge: Cambridge University Press.

Weinberg, S. (1989). "The cosmological constant problem", *Reviews of Modern Physics*, 61, 1-23.

Wolchover, N. (2015). "A fight for the soul of science", *Quanta Magazine*, issue of 16 December, https://www.quantamagazine.org/20151216-physicists-and-philosophers-debate-the-boundaries-of-science/ (accessed 14 March 2018).

Yeo, R. (1988). "Genius, method and morality. Images of Newton in Britain 1760-1860", *Science in Context*, 2, pp. 257-284."

Zaanen, J., Y. Liu, Y.-W. Sun, and K. Schalm (2015). *Holographic duality in condensed matter physics*, Cambridge: Cambridge University Press.